\documentstyle[prl,aps,psfig,twocolumn]{revtex} 

\begin{document}
\draft
\twocolumn[\hsize\textwidth\columnwidth\hsize\csname
@twocolumnfalse\endcsname
\title{Damaging and Cracks in Thin Mud Layers.}
\author{Raffaele Cafiero$^1$, Guido Caldarelli$^2$, Andrea Gabrielli$^3$} 
\address{$^1$ PMMH Ecole Sup. de Physique et de Chimie Industrielles 
(ESPCI), 10, rue Vauquelin-75231 Paris Cedex 05 France}
\address{$^2$ INFM Sezione di Roma1, Dipartimento di Fisica,
Universit\`a di Roma ``La Sapienza'', P.le
A. Moro 2, I-00185 Roma, Italy} 
\address{$^3$ Laboratoire de Physique de la Mati\`ere Condens\'ee,
Ecole Polytechnique, 91128-Palaiseau Cedex, France}
\date{\today}
\maketitle

\begin{abstract}
We present a detailed study of a two-dimensional minimal 
lattice model for the description of
mud cracking in the limit of extremely thin layers.
In this model each bond of the lattice is assigned to a (quenched) 
breaking threshold. Fractures proceed through the selection of
the part of the material with the smallest breaking threshold.
A local damaging rule is also implemented, by using two different types 
of weakening of the neighboring sites, corresponding 
to different physical situations. Some analytical results are derived through a 
probabilistic approach known as Run Time 
Statistics. In particular, we find 
that the total time to break down the sample grows with the dimension
$L$ of the lattice as $L^2$ even though the percolating cluster 
has a non trivial fractal dimension. Furthermore, a formula for
 the mean weakening in time of the whole sample is obtained. 
\end{abstract}
]
\narrowtext
\section{Introduction}
In this paper we present a careful and detailed study 
of a minimal fracture model
that has been introduced at the aim of describing the main features of 
paint dessication-like phenomena\cite{GCC}.
The purpose of this work is to focus on the 
statistical properties of these phenomena
on the basis of a recent experimental work\cite{mudcracks}. 
Following the results of this work, we assumed that
the main source of stress is given by the {\em local} 
friction between the layer of material and the bottom surface of the container. 
Moreover, it has been noticed that the characteristic size of crack 
patterns varies linearly with the layer thickness.  
In the limit of zero thickness {\em crack patterns lose 
their polygonal structure} (the characteristic size of the polygons is zero) 
and become {\em branched fractals}.

In order to model this behavior, a minimal automaton lattice model, 
inspired by Invasion Percolation\cite{IP} 
and by the vectorial and scalar models described in \cite{CDP,ZVS}, 
has been recently introduced by the authors \cite{GCC}. Here, we present
 an extended report of the study described in \cite{GCC}, with a detailed 
description of the analytical calculations a new numerical 
and theoretical results.
All the models for {\em quasi-static} fractures, describe crack 
evolution through a non-local Laplacian field (electric field, electric 
current) acting on a solid network of bonds or sites \cite{ZVS}. 
In others the stress field evolves by keeping minimum the energy of the system.
In such a case the components of the obtained vectorial equations are 
similar to the equations describing the action of a Laplacian field \cite{CDP}.
In this model,instead, no explicit field is present. 
The effect of the stress is 
represented by an extremal breaking rule and {\em local} 
random breaking thresholds: 
at each time step, the bond with the smallest 
threshold is removed from the lattice. 
Short ranged correlations are introduced through a damaging of the non-broken  
nearest neighbors bonds of the just removed bond. 
According to the kind of fracture we deal with, one can introduce
different types of damaging. In this paper 
two limiting cases are studied.
This model is inspired  by the above cited experimental observations
\cite{mudcracks} that, in 
an extremely thin layer of mud or paint, the only source of stress 
is the {\em local} friction with the container. 
Moreover, since the drying mud is a 
mixture of a liquid and a solid (usually amorphous) phase, no long range 
stress relaxation is present, although the growing crack can 
affect the properties of the medium in its neighborhood. 
Some important physical properties of the model are explained by using an 
approach based on the Run Time Statistics (RTS) scheme \cite{RTS}. 
In particular, we are able to compute 
some relevant quantities, such as the evolution of the breaking probability,
and of the probability distribution of breaking thresholds.

The paper is so organized. In Sec. II the model is described. In Sect. III 
the results of numerical simulations are presented. In Sect. IV the model
is studied 
analytically and theoretical and numerical results are compared.

\section{The Model}
A square lattice is considered and a 
quenched random variable $x_i$ is assigned to each bond $i$. 
The $x_i$'s are independently extracted from an
 uniform probability density between $0$ and $1$.
At each time-step $t$, the unbroken bond in the lattice with the lowest 
value of the variable is broken (removed). 
Then damage (weakening) is applied, in a way explained below, to the 
unbroken nearest neighbors (n.n.) of the just removed bond. 
After having introduced the damaging, the breaking and damaging steps 
are repeated until a connected, percolating, 
subset (infinite cluster) of removed bonds appears, dividing
 the system into two disconnected parts.

Before explaining the definition of damaging, it is necessary 
to introduce some notations. 
The set of broken bonds up to time $t$ is indicated with $C_t$,
and the set of non-broken bonds with $\partial C_t$.
The number of bonds belonging to $C_t$ is $\|C_t\|=t$, 
while $\|\partial C_t\|=N-t$ (where $N$ is the total 
number of bonds in the lattice), 
in fact $\partial C_t$
is composed by the whole lattice minus the bonds in $C_t$.
The definition of $C_t$ is independent of the model. 
That of $\partial C_t$, instead, can differ 
a lot from a model to another; for instance in Invasion Percolation (IP) it
is simply given by the set of nearest neighbors of the bonds in $C_t$. 

Two different 
kinds of weakening of the unbroken neighbors are studied:  
Either by direct weakening or by re-distribution of the ``stress''.
In the first case (rule {\bf 1}), the unbroken n.n. are 
weakened, by extracting 
a new threshold $x'_i$ between $0$ and the former value $x_i$. 
In this case an average weakening 
of one half of the former value at time is obtained.
In the second case (rule {\bf 2}), instead, each neighbor has a threshold 
weakened by a fraction of the threshold of the bond just removed.
Both cases mimic the damaging produced by the enhancement 
of the stress nearby crack tips: 
the first case refers to a situation where stochasticity 
(thermal fluctuations) is important
in the determination of the new thresholds 
\cite{HR}, the second case refers to a deterministic 
effect around the crack tip. 
From the point of view of mud cracking, the 
two-dimensional lattice represents a very thin layer of mud (or paint), and 
quenched disorder {\em accounts for local stress induced by inhomogeneous 
desiccation of the sample}. 
Since the evolution of cracks in mud desiccation 
is assumed to be a slow process, 
the dynamics is assumed to be {\em quasi-static}, i.e. one microscopical
breaking with relative damaging for each time-step.
Some authors  correctly point out that otherwise 
time-dependent effects and a non-equilibrium dynamics are relevant 
in crack propagation \cite{sornette}.

In this model the explicit presence of an external field
(applied stress) and of the response of the material
(strain of bonds) have been eliminated. The only quantity 
present is the breaking threshold, the
dynamics of which is chosen to reproduce the evolution 
of cracks. This simulates 
the presence of a local stress field, acting not on the boundaries but directly
on each bond. Our assumption is based on the experimental results in Ref. 
\cite{mudcracks}, where, as the mud layer becomes thinner, only the 
inhomogeneities drive the nucleation of cracks. Furthermore, the hypothesis of
crack developing under the same state of strain not only is usually applied
in the presence of thermal gradients \cite{mud0}, but is also commonly reported
in experiments of loading of softened material \cite{mud1,mud2,mud3}.
Hence, such a model is 
particularly suitable to describe, for example, paint drying, where the stress
applied to the painted surface depends on the local action of external
conditions (density gradient in the paint). Moreover, its simplicity 
allows us to study analytically its properties, which is a non common 
feature for fracture models.

\section{Numerical Results}
Numerical simulations, with cylindrical symmetry 
(periodic boundary conditions in the horizontal direction)  
for various system sizes $L$ have been performed. The dynamics stops 
as soon as a crack spanning the system in the vertical 
direction appears. Both damaging rules are implemented, 
and they are discussed in parallel.
Despite the simplicity of the dynamical rules, 
the results are rather interesting. 
We have computed the fractal dimension of the percolating cluster, the 
distribution of the size of clusters of broken bonds, 
the avalanche size-distribution (in order to check if long range temporal 
correlations are present), and the probability distribution 
of the breaking thresholds at the percolation time.
An avalanche can be defined as an ensemble of causally and geometrically
connected breakdowns (see below for a rigorous definition). Under this respect 
the size-distribution of such avalanches represents the probability of a 
large or small response of the system to an external solicitation.
For example a power law distribution represents a critical state of
the system where the response has not a characteristic size.

The fractal dimension $D_f$ of the percolating cluster is 
computed using the box-counting method. 
The analysis is restricted to the spanning cluster to 
reduce the finite size effects present for the smaller clusters. 
The results of the box-counting analysis are reported in Table
 \ref{table1} for the different sizes and for the two damaging rules. 
The values of $D_f$ for the two damaging rules coincide within the error bars.

\begin{figure}
\centerline{
\psfig{file=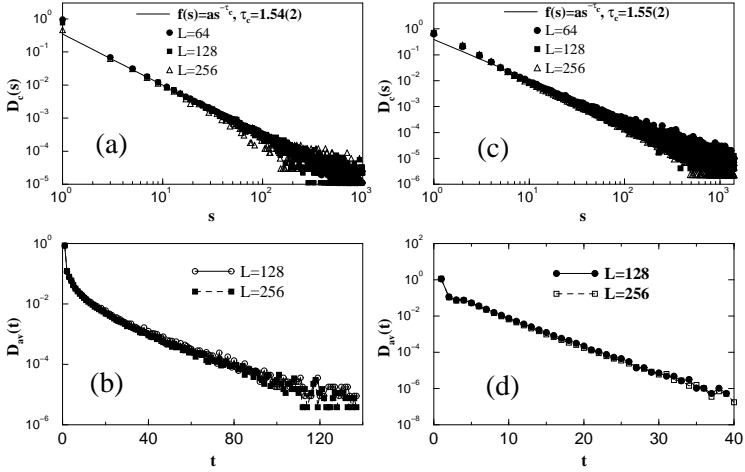,height=6.5cm,angle=-90}
}
\caption{(a): Probability distribution ($log_{10}$-$log_{10}$ plot) 
$D_c(s)$ of the cluster size, for $L=64,128,256$. 
(b): Avalanche size distribution (linear-$log_{10}$ plot) $D_{av}(t)$
for $L=128,256$. 
(c) and (d): The same quantities for the weakening rule {\bf 2}.}
\label{fig1}
\end{figure}

The connected clusters of broken bonds are identified 
with a standard cluster counting procedure, based on the 
Hoshen-Kopelman algorithm \cite{hoshen}.
Also the distribution of finite clusters is nontrivial,  
showing a clear power law with exponent $\tau_c=1.54(2)$ 
(see Fig.~\ref{fig1}(a)) for rule {\bf 1} and $\tau_c=1.57(3)$ 
for rule {\bf 2}.
The plots labelled with (b) in Fig.~\ref{fig1} refer to the avalanche
size distribution.
This quantity is interesting with respect to recent experiments 
\cite{PPVAC} and models\cite{CDP,ZVS} where a power law behavior of
the acoustic emission has been related to Self-Organized Criticality
(SOC)\cite{BTW}. The presence of a SOC-like behavior 
would mean that the dynamics 
of fractures itself leads the sample to a steady state where small 
variation of the external field can trigger reaction at any length-scale.
In particular the external field in this case is the applied stress, and the
response of the sample can be considered as the energy released (acoustic 
emission) 
by one avalanche of cracks, where avalanche means a 
causally and geometrically connected 
series of breakdowns.
In this oversimplified model the external stress can be considered constant, 
since the only change after any single breakdown is due the damaging of the 
n.n..
Consequently, in this work
the size of an avalanche is monitored as a measure for the acoustic emission.
An avalanche can be defined as follows.  
Let us suppose that a bond $i$ grows (i.e. it is broken) 
at time $t$; this is the {\em initiator} 
of an avalanche, which is defined as the set of events geometrically 
and causally connected to the initial one (bond $i$). ``Causal" 
connection refers to the weakening following any bond breaking.
In particular, when bond $i$ grows at time $t$, the avalanche goes on at time 
$t+1$ if a unbroken first neighbor bond $j$
of $i$ is removed. 
At time $t+2$ the avalanche goes on if a bond $k$ grows where $k$ is a 
unbroken first neighbor of $i$ or $j$ and so on.
A linear-log plot of the probability 
distribution of avalanche size, versus sample 
size $L$ is shown in Fig.~\ref{fig1}b.  
After a power law transient, an exponential distribution is reached, 
indicating that a characteristic size exists for the avalanches. 
One can note that both for weakening rule {\bf 1} and weakening rule
{\bf 2} simulations give qualitatively similar results, although for
 rule {\bf 2} the characteristic time of avalanches is smaller. 
This is easily explained, since the damaging rule {\bf 2} is 
less strong than rule {\bf 1}, and, consequently, the causal connection 
between subsequent breaking events is weaker.

This result for avalanches is similar to those obtained for a scalar 
model of Dielectric Breakdown, but differs from the avalanche 
behavior in models of fracture \cite{CDP,ZVS}.
The explanation of this behavior is motivated by two arguments. 
Firstly, in the present 
definition of an avalanche the threshold
is changed only for the n.n.. This introduces a typical length 
scale, while other definitions consider as the threshold the ratio between 
local field and resistivity, thus
giving the possibility of large scale correlations. 
Secondly, in this model broken bonds {\em are removed from the system}. 
This represents a substantial difference with many SOC models 
with quenched disorder
presented in the literature.
For example, in a simple toy model of SOC due to Bak and Sneppen\cite{BS}
(where a similar refresh of thresholds is present) 
the dynamics produces clear power laws in the 
avalanche distribution. 
There, each site (species) deleted is replaced by a new one
and is not definitively removed. In our model, instead, the 
number of candidates $\partial C_t$ to be broken at each time-step decreases 
in time.
This is a crucial point, since indeed power law behavior in the presence of a 
scalar field seems to be related to a ``reconstructing rule'' that allows one 
to deal with a system where removed bonds are replaced by new ones. 
Therefore, only in the case of plastic deformation, one is in the presence 
of a steady state, as correctly pointed out by Ref.\cite{ZVS}.

Therefore, the fractal dimension, the cluster 
size distribution and, to some extent, the avalanche size distribution 
seem to be universal with respect to the two different local damaging rules. 
In the next section the study will focus on some quantities 
which, instead, are not universal and reproduce the evolution of the 
mechanical properties of the material during the fracturing dynamics. 
These quantities are the average probability density of breaking thresholds, 
or histogram, $\phi_t(x)$, and as a by-product the mean breaking threshold 
$\langle x \rangle(t)$, which expresses the average resistance 
to breaking, or rigidity, of the system at time $t$. 
These quantities will be studied both numerically, and analytically, 
by using a probabilistic tool called Run Time Statistics (RTS) \cite{RTS}.

\section{RTS Derivation of the average weakening of the material}
As seen above, the evolution of the crack is described by a quasi-static 
extremal dynamics in a medium with quenched disorder.
The most important question for a theoretical comprehension of the model
is: which is the source of the 
spatio-temporal correlations developed by the dynamics?
As pointed in \cite{RTS} in relation to Invasion Percolation (IP), the source
can be found in the memory effects developed by the evolution of 
the dynamics itself via an interplay between dynamical rules and
 quenched disorder.

This can be simply realized observing that the knowledge of the
 growth history up to a time $t$,
provides information about the probability distribution and the correlations 
of the random bond-thresholds.
This information has to be added to the original 
information that the thresholds are
independently extracted from the uniform probability density in the interval $[0,1]$.
Moreover this information influences the probabilities of the 
different possible continuations of the dynamics for larger time.
This memory effects can be studied using carefully the notion
of conditional probability.
This kind of approach to growth dynamics with quenched disorder
has been developed in \cite{RTS,RTS-rig}, with particular reference to IP.
This peculiar probabilistic algorithm is called 
{\em Run Time Statistics} (RTS).
A particular modification of this tool is presented here taking into account
the damaging mechanism, which is not present in IP-like models.
Finally, RTS is used in order to predict analytically some relevant 
quantities as the evolution of both the average 
probability density of breaking thresholds 
of unbroken bonds and of the mean resistance to breakdown $x(t)$ 
of the material.

Here we provide directly the final RTS formulas 
together with a brief sketch about their meaning. 
A detailed derivation of the analytical results of this section 
is given in Appendix $A$.
The RTS approach permits mainly to answer the following two questions, 
once given a certain time-ordered geometrical path followed by the dynamics
up to time $t$: 
\begin{enumerate}
\item which is the {\em effective} probability density function
of the variables $x_i$ of the lattice conditioned to 
the knowledge of this fixed past dynamical history;
\item which is the conditional probability of any further growth event
at the next time-step.
\end{enumerate} 

In order to introduce operative formulas, let 
us think to know the ``one-bond'' 
effective probability density functions $p_{i,t}(x)$ 
(conditioned to the past dynamical 
history) for each non-broken bond $i$. As it is 
clarified in Appendix A, this ``one-bond''
formulation of RTS is an approximate of the rigorous one. 
However, as shown in
\cite{RTS-corr}, it is a good approximation when the number 
of random numbers is large
(as in this case).

First of all one can write \cite{RTS}
the breaking probability $\mu_{i,t}$ 
for each bond $i$ at that time-step:
\begin{equation}
\mu_{i,t}=\int_0^1 dx\,p_{i,t}(x) \left[ \prod_{k(\neq i)}^{\partial C_t} 
\int_x^1 dy \, p_{k,t}(y) \right]\,,
\label{mu}
\end{equation}
where $\partial C_t$ is the whole set of unbroken bonds. 
Note that at time $t$ the number of bonds in $\partial C_t$ is 
$(2L^2 - t)$, i.e., the total number $2L^2$ of bonds in a square lattice 
of side $L$ minus the number of broken bonds before time $t$.
Eq.~(\ref{mu})  expresses nothing else than the effective probability 
that $x_i$ is the minimum in the set $\partial C_t$ conditioned 
to the past history.  
The next important step is to update each $p_{j,t}(x)$ by 
conditioning them to this latest growth event. In this way one obtains the 
$p_{j,t+1}(x)$'s conditioned
to the history up to the time-step $t+1$. 
The effective probability density at time $t+1$ of the 
latest grown bond $i$ is usually called $m_{i,t+1}(x)$, 
in order to distinguish it from the densities of still unbroken 
bonds. 
It is given by
\begin{equation}
m_{i,t+1}(x)=\frac{1}{\mu_{i,t}} p_{i,t}(x)\left[ 
\prod_{k(\neq i)}^{\partial C_t} \int_x^1 
dy\,p_{k,t}(y) \right] .
\label{m}
\end{equation}
Eq.~(\ref{m}) (multiplied by $dx$)  gives the ``effective'' 
probability that $x\le x_i\le x+dx$,
conditioned to the past fixed dynamical history (time-ordered path) 
up to time $t+1$: 
the ``memory'' of the history up to time $t$ is ``recorded'' in the 
set of functions 
$p_{k,t}(x)$, where $k$ runs over all the bond belonging to $\partial C_t$,
while the last step is recorded in the particular functional relationship
between $m_{i,t+1}(x)$ and the set $\left\{p_{k,t}(x)\right\}$ itself.
This relationship is imposed by the order relation among the interface 
variable $x_k$,
i.e. by the fact that $x_i$ is the minimum in $\partial C_t$.
Note that, once a bond is broken, it does not participate anymore 
to the dynamics.
For this reason, the ``effective'' probability density function of its threshold 
does not change anymore in time and is given definitely by Eq.~(\ref{m}).

For the remaining bonds one has to distinguish among the unbroken bonds
far away from the bond $i$ and the unbroken nearest neighbors
 bonds, which will be weakened by the growth of bond $i$. 
The updating rules, for the two different
mechanisms of damaging, differ only for this last set of bonds.
For the non-weakened bonds, one has in both cases the 
following updating equation:
\begin{equation}
p_{j,t+1}(x)=\frac{1}{\mu_{i,t}} 
p_{j,t}(x)\!\!\int_{0}^{x}\!\! dy \,p_{i,t}(y) 
\!\!\left[ \prod_{k(\neq i,j)}^{\partial C_t}\!\!
\int_y^1 \!\!dz \, p_{k,t}(z) \right] 
\label{p1}
\end{equation}

The updating equations for the weakened bonds are instead the following:\\
{\bf (1)} For the damaging mechanism {\bf 1}:
\begin{eqnarray}
&&p_{j,t+1}(x)=\frac{1}
{\mu_{i,t}}\int_{0}^{1}dy\, \frac{1}{y} \theta(y-x)
p_{j,t}(y)\times\nonumber\\
&&\times\int_{0}^{y} dz \, p_{i,t}(z)
\left[ \prod_{k(\neq i,j)}^{\partial C_t}
\int_z^1 du \, p_{k,t}(u) \right]\,.
\label{p21}
\end{eqnarray}
{\bf (2)} For the damaging mechanism {\bf 2} (see Appendix A):
\begin{eqnarray}
&&p_{j,t+1}(x)=\nonumber \\
&&=\frac{1}{\mu_{i,t}}\!\int_0^1\!\! dy\!
\left[\!\prod_{k(\neq i,j)}^{\partial C_t}\!\int_y^1\!\! dz\, p_{k,t}(z)\right]
p_{i,t}(y)\,p_{j,t}\!\!\left(x+{y\over n_{i,t}}\right)\times
\nonumber \\
&&\times\theta\left(\frac{n_{i,t}}{n_{i,t}-1}x-y\right)
\theta\left(n_{i,t}(1-x)-y\right)
\label{p22}
\end{eqnarray}
Note that the main difference between Eq.~(\ref{p21}) and 
Eq.~(\ref{p22}) is due to
the fact that the number of n.n. $n_{i,t}$ of the bond $i$ at time $t$ 
appears explicitly only in the 
latter, i.e. only in the second model (rule {\bf 2}) the 
damaging is an explicit function of the geometry, 
while in the former (rule {\bf 1})
the damaging is a ``one-bond'' process.

Eqs.~(\ref{mu}-\ref{p1}) coincide with the 
ones introduced for the RTS approach
to IP (apart from the different definition of the 
growth interface $\partial C_t$). Eqs. (\ref{p21}), (\ref{p22}), 
instead, are new and account for the n.n. weakening.
Eqs.~(\ref{mu}-\ref{p22})
allow one to study the extremal 
deterministic dynamics as a kind of stochastic
process with memory. In particular, $\mu_{i,t}$ can be used to evaluate 
systematically the statistical weight of a fixed 
time-ordered growth path, while 
the $p_{j,t}(x)$'s store information about the growth history.

A very important quantity to characterize the properties of the dynamics
is the empirical distribution (or histogram) of unbroken thresholds.
This quantity is defined as:
\begin{equation}
h_t(x)=\sum_{j \in \partial C_t} p_{j,t}(x)
\end{equation}
where, $h_t(x)dx$ is the number of non-broken bonds between $x$ and  $x+dx$
at time $t$, conditioned to the past dynamical history. 

Considering the effect of the growth of bond $i$ at time $t$ on this
quantity, one gets
\begin{equation}
\!\!h_{t\!+\!1}(x)\!=\!h_t(x)\!-\!m_{i,t\!+\!1}(x)\!-\!\!\sum_{j(i)}
p_{j,t}(\!x)\! +\!\sum_{j(i)}p_{j,t+1}\!(\!x)
\label{found}
\end{equation}
where $j(i)$ indicates the sum over the $n_{i,t}$ unbroken n.n. of $i$.
Moreover, $m_{i,t+1}(x)$ and $p_{j,t+1}(x)$ are given respectively by 
Eq.~(\ref{m}) and Eqs.~(\ref{p1}), (\ref{p21}) 
( (\ref{p22}) for rule {\bf 2}) .
Being the histogram an almost self-averaging quantity of the model
in the large time limit, one can 
evaluate its shape in the ``typical'' realization of the dynamics
taking the average over all the possible histories up to time $t+1$.
The notation $\left\langle...\right\rangle$ is introduced to indicate
for this average. The l.h.s. of 
Eq.~(\ref{found}) can be computed as
\begin{equation}
\left\langle h_{t+1}(x)\right\rangle =\!||\partial C_{t+1}||\phi_{t+1}(x)\!=
\![N\!-\!(\!t\!+\!1\!)]\phi_{t+1}(x), 
\label{triv}
\end{equation}
where $N=2L^2$ is the total number of bonds in the lattice and 
$\phi_t(x)$ represents the average thresholds density function over
the unbroken bonds at time $t$ (normalized to $1$), i.e.
$\phi_t(x)=p_{k,t}(x)$ where $k$ is a generic interface bond.
For the r.h.s. of Eq.~(\ref{found}) the main difficulty arises in 
the evaluation of $\left\langle m_{i,t+1}\right\rangle$ and 
$\left\langle\sum_{j(i)} p_{j,t+1}(x)\right\rangle$.
Following \cite{RTS}, one can write 
\begin{equation}
\left\langle m_{i,t+1}\right\rangle\simeq (N-t)\phi_t(x)
\left[1-\int_0^x dy\phi_t(y)\right]^{N-t-1}
\label{mind}
\end{equation}
In obtaining Eq.~(\ref{mind}), we used the definition of
$\phi_t(x)$ and the following approximation:
\begin{equation}
\left<\prod_{k\in\partial C_t}p_{k,t}(x_k)\right>=\prod_{k\in\partial C_t}
\left<p_{k,t}(x_k)\right>=
\prod_{k\in\partial C_t} \phi_t(x_k)\,.
\label{appr}
\end{equation}
Using again the definition of $\phi_t(x)$,
one gets
\begin{equation}
\left\langle\sum_{j(i)} p_{j,t}(x)\right\rangle=n_t\phi_t(x),
\end{equation}
where $n_t=\left<n_{i,t}\right>$.
Using Eq.~(\ref{p21}), corresponding to the weakening
 rule {\bf 1}, and the approximations given by
Eq.~(\ref{appr}), one has 
\begin{eqnarray}
&&\left\langle\sum_{j(i)} p_{j,t+1}(x)\right\rangle
=\frac{n_t(N\!-\!t)}{N\!-\!t\!-\!1}\int_x^1 \!\!\!dy\,\frac{\phi_t(y)}{y}
\times\nonumber\\
&&\times \left\{1\!-\left[1-\int_0^y dz\phi_t(z)\right]^{N-t-1}\right\}.
\label{MFweak1}
\end{eqnarray}

The equation for the $\phi_{t+1}(x)$ for rule {\bf 1} will finally read:
\begin{eqnarray}
&&\phi_{t+1}(x)=\frac{N-t-n_t}{N-t-1}
\phi_t(x)+\nonumber\\
&&-\!\frac{\!N\!-\!t\!}{N\!-\!t\!-\!1}\phi_t(x) 
\left[1-\int_0^x dy\phi_t(y)\right]^{N-t-1}+ \nonumber \\
&&+n_t\frac{\!N\!-\!t\!}{(\!N\!-\!t\!-\!1\!)^2}\int_x^1dy
\frac{\phi_t(y)}{y}\times\nonumber\\
&&\times \left\{1-\left[1-\int_0^y dz \phi_t(z)\right]^{N-t-1}\right\}
\label{equazione}
\end{eqnarray}

\begin{figure}
\centerline{\psfig{file=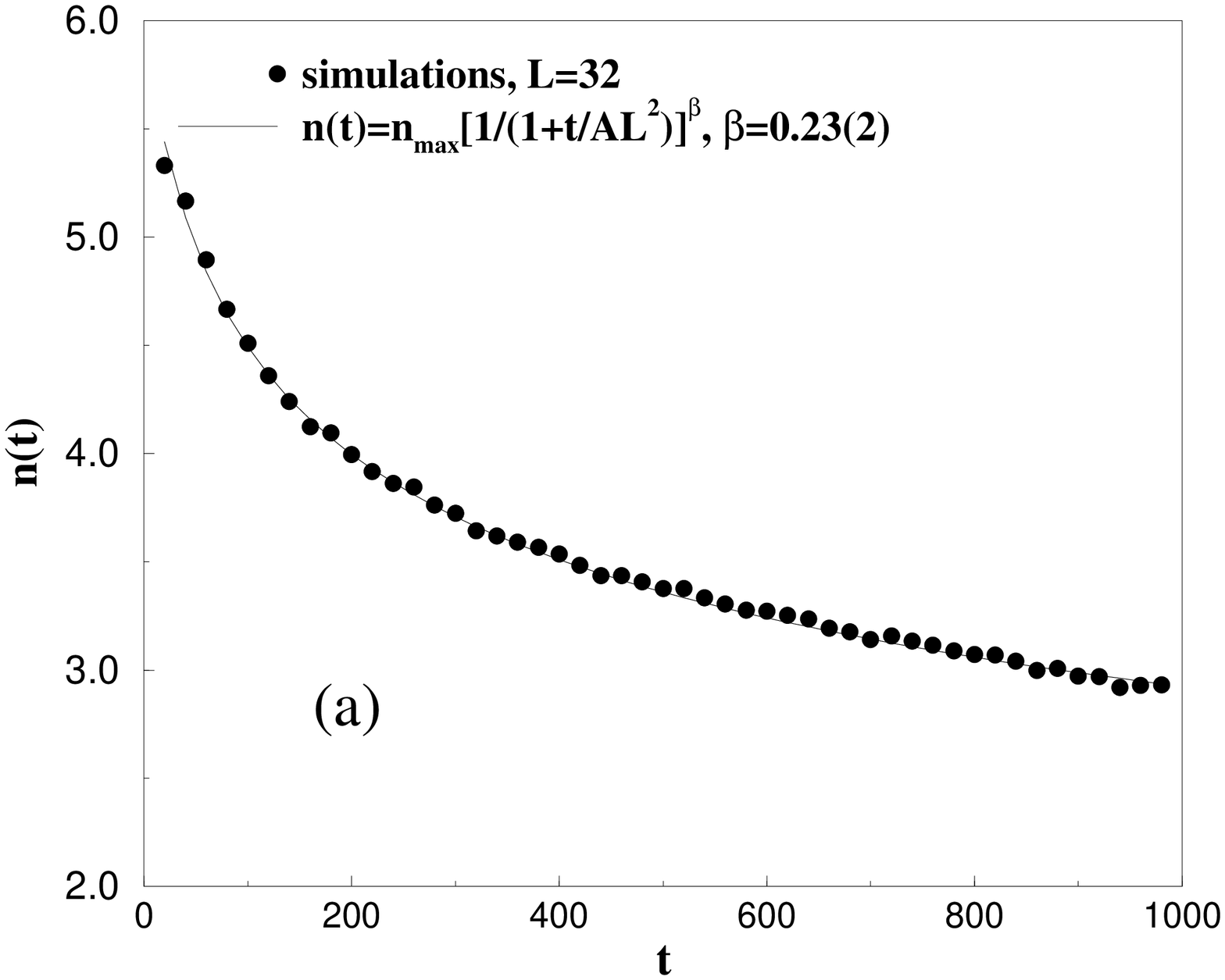,height=5.0cm,width=6.25cm,angle=-90}}
\centerline{\psfig{file=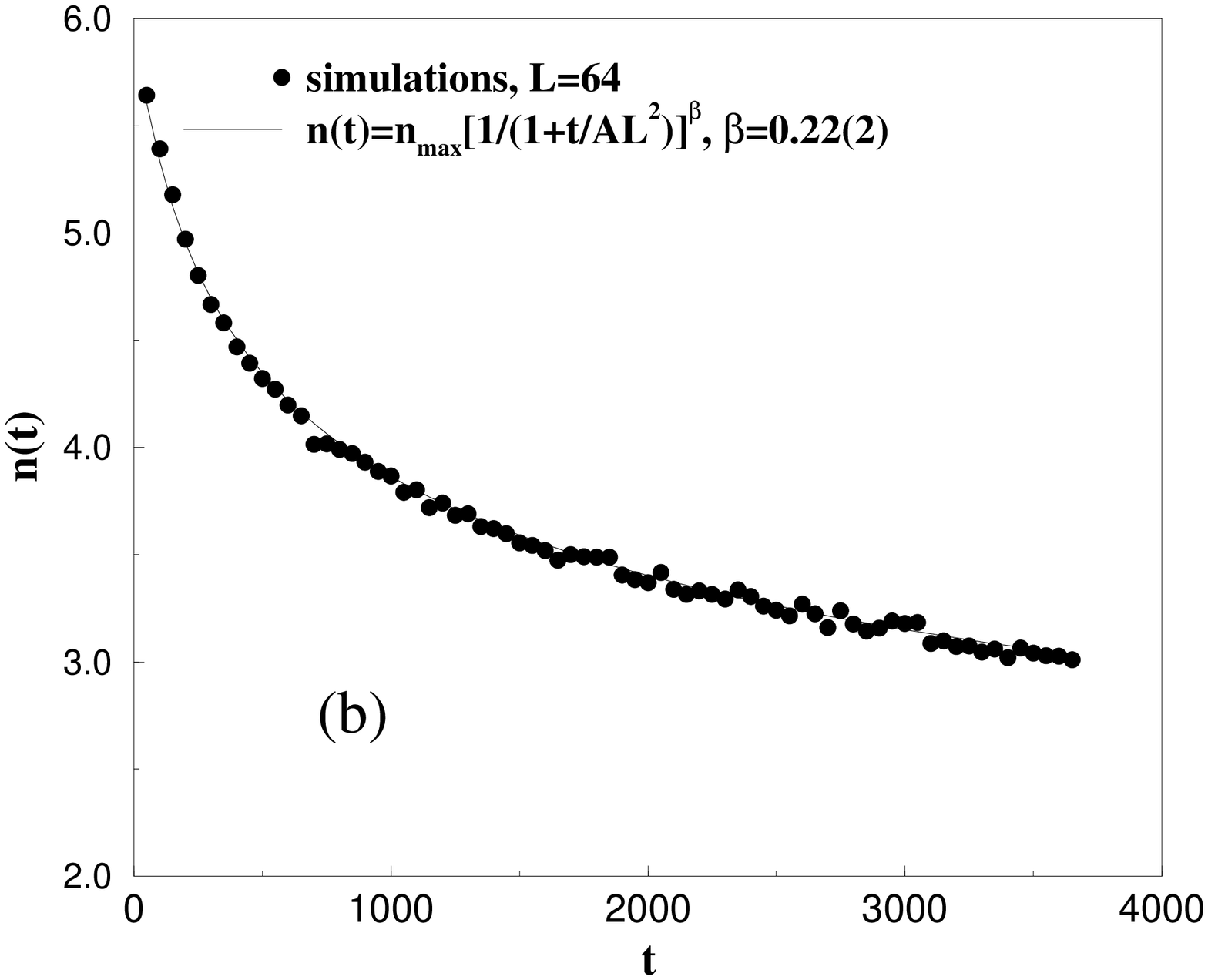,height=5.0cm,width=6.25cm,angle=-90}}
\caption{fit of $n_t(L)$ with the scaling form \ref{nt} 
(weakening rule {\bf 1}) for $L=32$ (a) and $L=64$ (b).}
\label{fig2}
\end{figure}

Note that even at percolation time $N-t$ is a large number. For this reason 
terms  in Eq.~(\ref{equazione})
containing the term $\left[1-\int_0^x dy\phi_t(y)\right]^{N-t-1}$ are 
negligible for $x$ such that $\int_0^x dy\phi_t(y)$ is finite 
(i.e. larger than $1/(N-t)$).
It is easy to show that the continuum limit of Eq.~(\ref{equazione}), for 
such values of $x$,
is invariant under the rescaling $L\rightarrow aL$ (i.e. $N \rightarrow 
a^2N)$ and $t\rightarrow a^2t$. 
This result is based on the assumption that:
\begin{equation}
n_t(L)=n_{a^2t}(aL)
\label{n}
\end{equation}
 The numerical simulations suggest the following 
scaling form for $n_t(L)$ (see Fig.~\ref{fig2}):
\begin{equation}
n_t(L)=n_{max}\left[\frac {1}{1+t/AL^2}\right]^\beta ,
\label{nt}
\end{equation}
where $\beta=0.23(2)$, $A=0.030(2)$ and $n_{max}=6$ 
is the lattice coordination number.
This form for $n_t$ satisfies Eq.~(\ref{n}).  

The study of the weakening rule {\bf 2} is quite
similar. Eqs.~(\ref{mu}-\ref{p1}) keep 
the same, while the conditioned probability density 
for the weakened bonds is given by Eq.~(\ref{p22}).

\begin{figure}
\centerline{\psfig{file=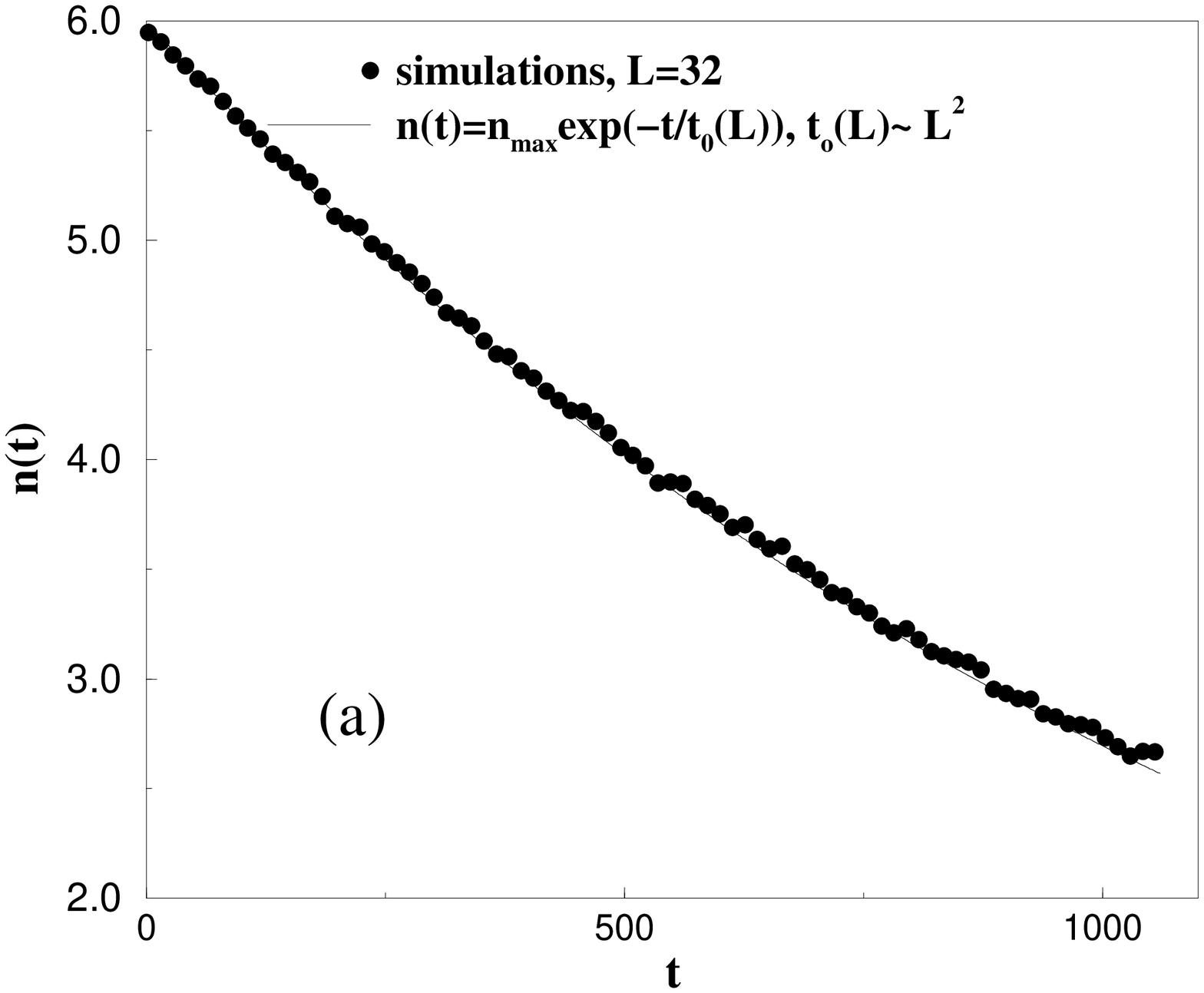,height=5.0cm,width=6.25cm,angle=-90}}
\centerline{\psfig{file=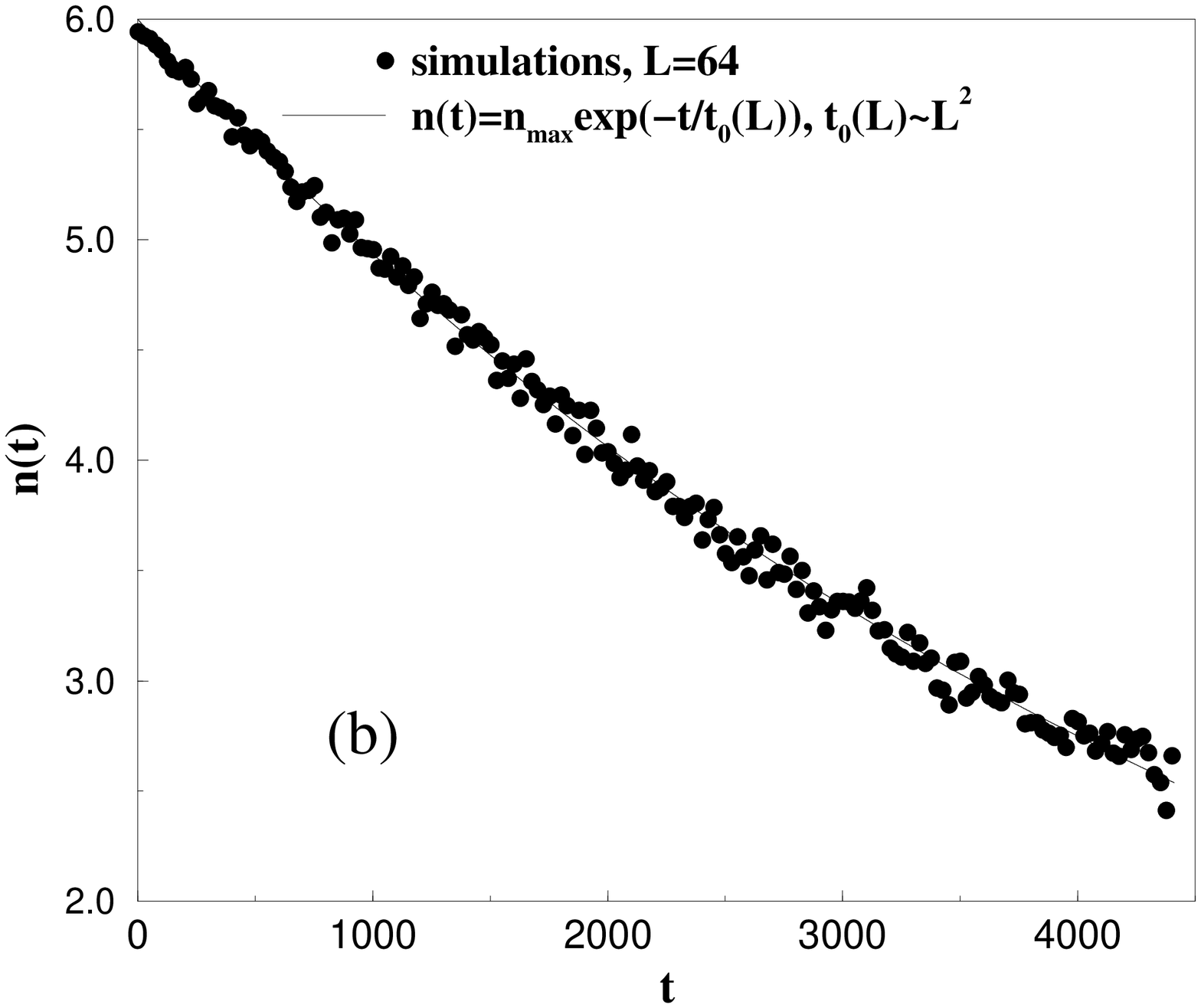,height=5.0cm,width=6.25cm,angle=-90}}
\caption{fit of $n_t(L)$ with the scaling form \ref{nt} 
(weakening rule {\bf 2}) for $L=32$ (a) and $L=64$ (b).}
\label{fig3}
\end{figure}

\begin{figure}
\centerline{
\psfig{file=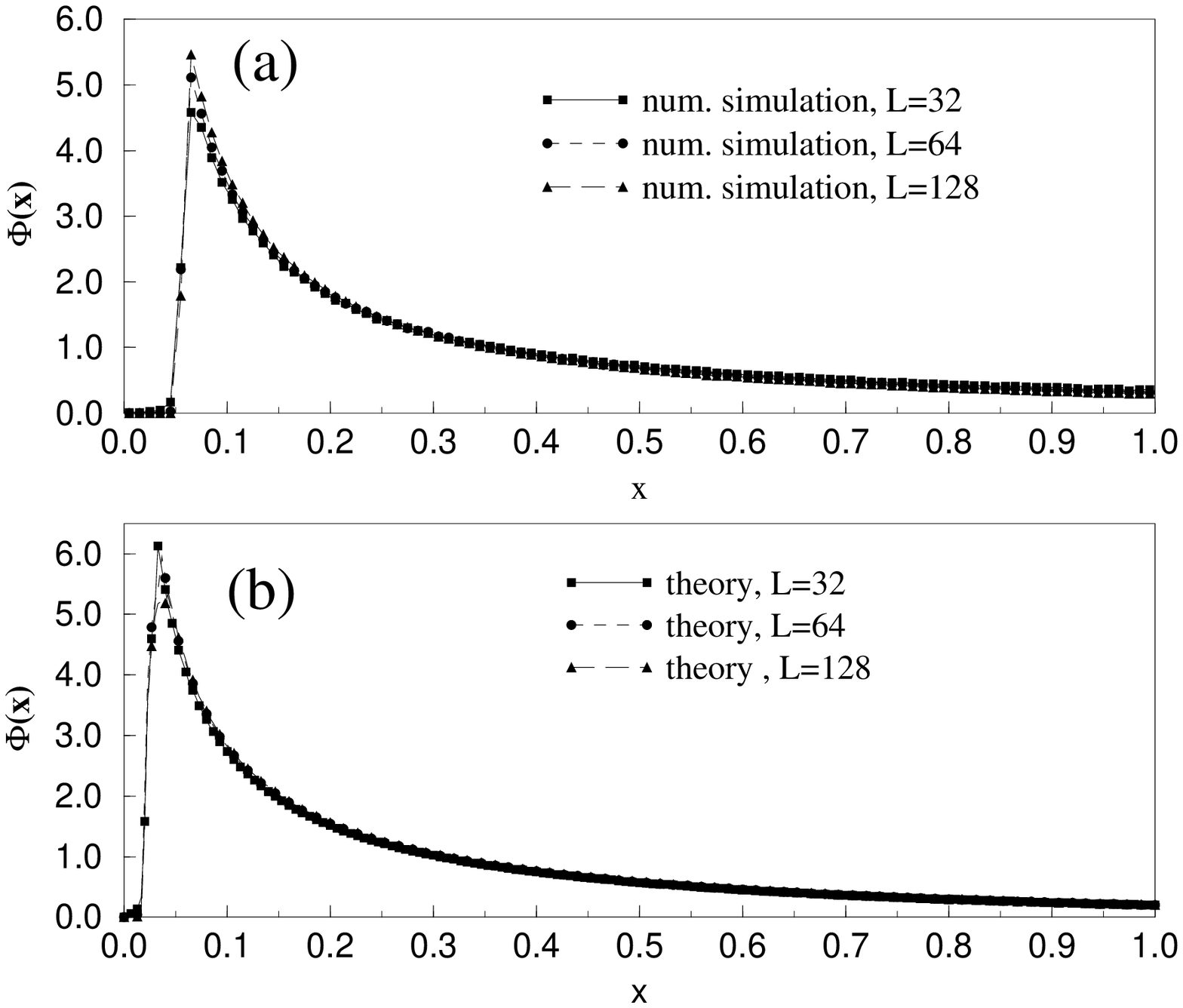,height=6.0cm,angle=-0}}
\caption{Solution of Eq.~\ref{equazione} for the histogram 
$\phi_t(x)$ at the spanning time (b), compared with simulations (a), 
for the weakening rule {\bf 1}.}
\label{fig4}
\end{figure}

\begin{figure}
\centerline{
\psfig{file=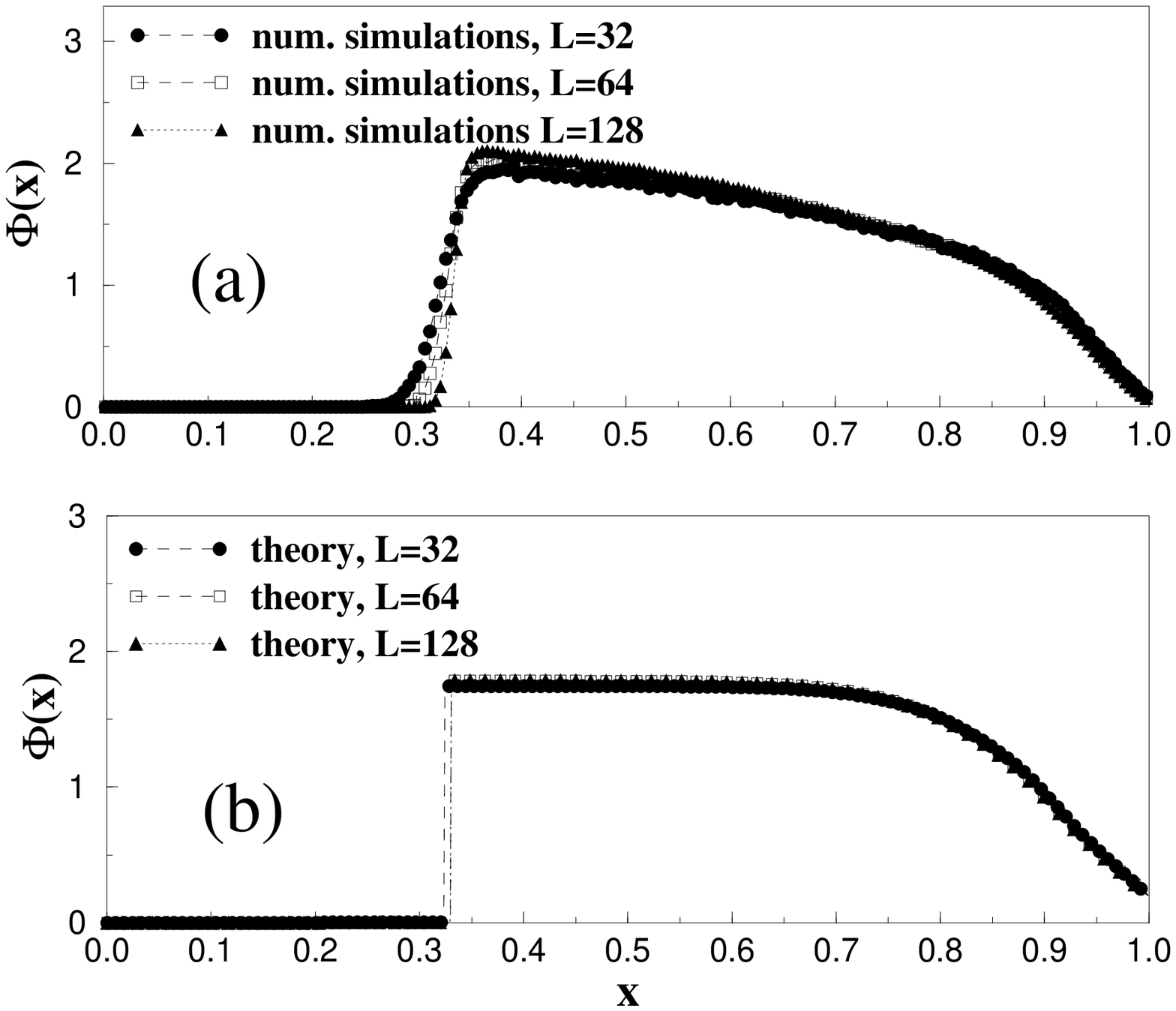,height=6.0cm,angle=-0}}
\caption{Solution of Eq.~\ref{equazione} for the histogram 
$\phi_t(x)$ at the spanning time (b), compared with simulations (a), 
for the weakening rule {\bf 2}.} 
\label{fig5}
\end{figure}

\begin{figure}
\centerline{\psfig{file=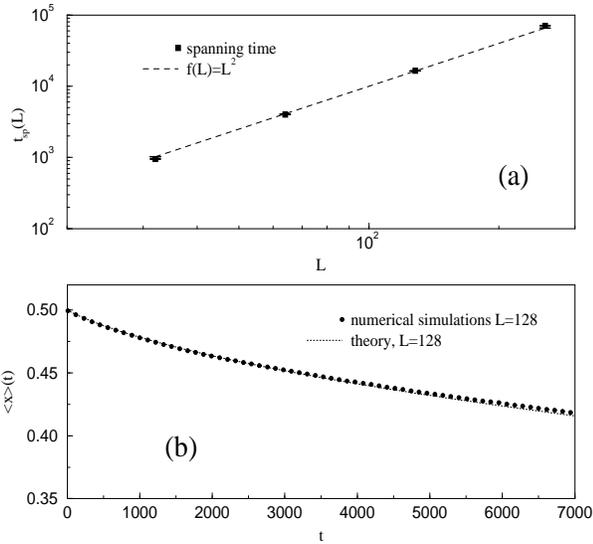,height=7.0cm,width=7.93cm,angle=0}}
\caption{(a) Spanning time versus system size $L$ for 
weakening rule {\bf 1}. One can see
 a nice agreement with the expected scaling law $t_{sp}(L)\propto L^2$.
(b) Solution of Eq.~\ref{xmed} compared with numerical simulations.}
\label{fig6}
\end{figure}

\begin{figure}
\centerline{\psfig{file=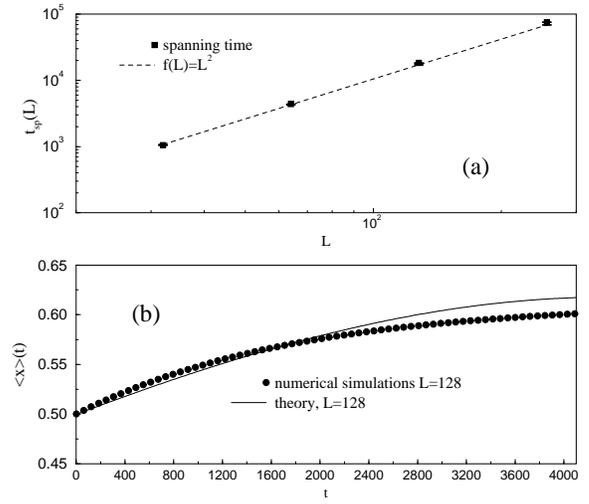,height=6.7cm,width=7.59cm,angle=-90}}
\caption{(a) Spanning time versus system size $L$ for 
weakening {\bf 2}. Also in this case,there is very good
 agreement with the expected scaling law $t_{sp}(L)\propto L^2$.
(b) Solution of Eq.~\ref{xmed2} compared with numerical simulations.}
\label{fig7}
\end{figure}

By following the same steps as above, the following equation 
for the $\phi_{t+1}(x)$ is obtained:
\begin{eqnarray}
&&\phi_{t+1}(x)=\frac{N-t-n_t}{N-t-1}\phi_t(x)+\nonumber \\
&&-\frac{N-t}{N-t-1}\phi_t(x)\left[1-\int_0^x dy\phi_t(y)\right]^{N-t-1}+
 \nonumber \\
&&+n_t \frac{N-t}{N-t-1}\int_0^1 dy
\left[1-\int_0^y dz\phi_t(z)\right]^{N\!-\!t\!-\!2}\phi_t(y)\times
\nonumber \\
&&\times
\phi_t\left(x+\frac{y}{n_t}\right)\theta\!\left(\frac{n_t}{n_t-1}x-y\right)
\!\theta(n_t(1-x)-y)
\label{equazione2}
\end{eqnarray}
All the assumptions we made for the case {\bf 1}, 
including the scaling {\em ansatz} given in Eq.~(\ref{n}), 
are valid for case {\bf 2}.
In particular, from numerical simulations, one can 
find the following behavior for $n_t(L)$ (see Fig.~\ref{fig3}) :
\begin{equation}
n_t(L)\simeq n_{max}\exp\left(-\frac{t}{A\,L^2}\right)\,.
\label{n2}
\end{equation}
The analytical study of both kinds of weakening allows
 to make three important predictions:
{\bf (1)} Firstly, we find both theoretically, from the numerical solution of 
Eqs.(\ref{equazione}, \ref{equazione2}), and from 
the numerical simulations of the model, 
a discontinuity in the histogram (see Fig.~\ref{fig4} 
and \ref{fig5}), indicating that the 
system evolves in such a way as to remove all bonds with threshold 
smaller than some critical value $x_c$.
{\bf (2)} Second, from the symmetry properties of 
Eqs.~(\ref{equazione},~\ref{equazione2}), 
we deduce that the number $t_{sp}(L)$ of broken bonds at the percolation time 
is proportional to $L^2$, even though the percolating cluster is fractal. 
This result, confirmed by numerical simulations (see Fig.~\ref{fig6}a, Fig.~
\ref{fig7}a), and compatible with the scaling function (\ref{nt}) 
for $n_t(L)$, is deduced supposing that at the percolation time the shape of 
the histogram is independent of $L$, an assumption which fits well 
with the numerical histogram (see Fig.~\ref{fig4}a and Fig.~\ref{fig5}a).
{\bf (3)} Finally, we present an approximated result for the dynamical
behavior of the average value (over the unbroken bonds) of the 
thresholds $\left\langle x\right\rangle (t)$.
This quantity can be seen as a characterization of the average resistance
of the material in time.

In order to find the evolution equation of $\left\langle x\right\rangle (t)$
for the damaging rule {\bf 1,} it is enough 
to multiply both sides of, respectively, 
Eq.~(\ref{equazione}) and Eq.~(\ref{equazione2}) for $x$ and
 integrate them in the whole interval $[0,1]$.
Then one finds:
\begin{eqnarray}
&&\langle{x}\rangle(t+1)=\left(1-\frac{n_t-2}{2(N-t-1)}\right)
\langle{x}\rangle(t)+\nonumber \\
&&-\frac{1+n_t/[2(N-t-1)]}{N-t-1} 
\int_0^1dx\left[1-\int_0^x dy\phi_t(y)\right]^{N-t}\,.
\label{xmed}
\end{eqnarray}
For the damaging rule {\bf 2}, the way to find 
the equation for $\left<x\right>$
is even simpler. In fact, it is enough to consider that at each
time step, the global effect on $\left<x\right>$ is 
equivalent to remove two bonds with 
the resistance equal to the minimal one at that time.
Therefore, one can write:
\begin{eqnarray}
&&\left<x\right>(t+1)=\left(1+\frac{1}{N-t-1}\right)\left<x\right>(t)+\nonumber\\
&&-\frac{2}{N-t-1}\int_0^1 dx\left[1-\int_0^x dy\phi_t(y)\right]^{N-t}\,.
\label{xmed2}
\end{eqnarray}

For rule {\bf 1}, it is simple to see, from Eq.~(\ref{xmed}),
that $\langle{x}\rangle(t+1) < \langle{x}\rangle(t)$ 
until $n_t>2$ (which is verified for all the times).
This means that on average the medium weakens during the evolution
even if the weakest bond is removed at any time step.
This is due to the fact that, in this case the weakening of the neighbors
of the weakest interface bond has a stronger effect on the material than the
removal of the weakest bond itself. 
For the rule {\bf 2}, instead, one finds that 
$\langle{x}\rangle(t+1) > \langle{x}\rangle(t)$, if $\left<x\right>(t)$
is larger than the double of the average minimal threshold, and,
due to the extremal nature of the dynamics,
this is verified always in the large $N$ limit, i.e. in the limit of a
large number of bonds in the interface at any time-step. 
This means that in this second case, damaging is not strong enough to allow a 
global weakening of the system, which becomes more and more rigid. 
This is reasonable since in rule {\bf 2} 
the stress on the weakest bond is redistributed to the nearest 
neighbors, and the total initial stress is conserved, 
while in rule {\bf 1} there is not total stress conservation.
In other words, in the model with rule {\bf 1} the damaging is a 
multiplicative effect, i.e. the damaging is proportional to the 
old threshold (which can be big), in the model with rule 
{\bf 2} the damaging is quite reduced by the 
fact that at each time-step it is proportional to the 
minimal threshold in the whole system.
In Figs.~\ref{fig6}b, \ref{fig7}b the time evolution of
 $\langle{x}\rangle(t)$ obtained from computer simulations is compared with 
the theoretical prediction. 
Our analytical results are in good agreement with numerical simulations. 
For rule {\bf 2}, numerical simulations of the histogram 
evidentiate a low $x$ tail below the critical threshold, 
which tends to disappear as the 
system size grows, and a non zero slope of the part just above 
the critical threshold. The first one is a clear finite size effect, which 
is less important in the simulations of rule {\bf 1}, because 
for rule {\bf 1} the critical threshold is very small. Of course, 
such a finite size effect is absent in the theoretical results, 
as in all mean field (MF) approaches. The second effect could be
 due to spatial correlation induced by 
the damaging rule {\bf 2}, which in the analytical approach 
are neglected. This second effect does not disappear 
as the system size grows. 
Consequently the agreement between the
 numerical simulations of $\langle{x}\rangle(t)$ 
and Eq.~(\ref{xmed2}) is less good 
than for rule {\bf 1}. The numerical  
 $\langle{x}\rangle(t)$, mainly because of the nonzero negative slope
of $\phi(x)$above $x_c$, is a bit smaller than the 
theoretical prediction.  

With respect to real fracturing processes the behavior of the 
average resistance 
$\left<x\right>(t)$ obtained with rule {\bf 2} 
is more realistic, since in real materials one usually observes 
that the material during micro-cracks formation becomes 
more rigid, although more fragile, since the number 
of bond one has to break to have global breakdown 
becomes smaller and smaller. 
Moreover it is worth to note 
that, apart the shape of $\phi(x)$ and the behavior 
of $\langle{x}\rangle(t)$, all the other statistical 
properties of the system do not depend on the 
used weakening rule.

Finally, it is worth to point out that, to our knowledge, 
apart the qualitative results of \cite{mudcracks}, 
no quantitative experimental results are available. 
For example, a measurement of the fractal dimension of 
cracks or their size distribution would be extremely 
useful to further test the predictions of this model.
At the moment, this model seems able to capture, with 
its extremely simplified dynamics, some basic properties 
of fracturing processes.

In conclusion, we have presented a new model for fractures, which 
is useful in describing in a semi-quantitative way some basic mechanism in
 drying paint-like and mud-like cracking processes, 
for extremely thin samples. Due to its extreme simplicity,  
the model is particularly suitable for large scale simulations and takes into
account the damaging effects involved in fracture propagation. 
Even in this simple model we are able to analyze 
which conditions trigger SOC behavior in such systems. 
Furthermore, the change in the threshold distribution, induced by the 
damaging mechanism, allows us to write down explicitly the form 
of the breakdown probability for the bonds of the sample. Possible further research  
 could consist, for damaging rule {\bf 2}, in a more refined calculational scheme, in which
 two variable probability densities are also considered. This would be the first order
correction to our MF approach considering only one-variable distributions, and could allow
 us to take into account correlations induced by the damaging rule. Such a generalization
 of the RTS theory, formally discussed in \cite{RTS-rig}, is however technically very difficult.
Another research direction we are following is the application of real space techniques,
 combined with the RTS approach, to calculate the critical exponents of the model.   
The authors acknowledge the support of the EU grant Contract No. FMRXCT980183.

\appendix
\section{RTS for the damaged system}

In this appendix a simple explication of the RTS probabilistic
equations is provided.

First of all, one has to note that in this kind of models (as well as 
in Invasion Percolation) the initial
condition of the system is characterized by independent variables
(the breaking thresholds of the bonds) identically and uniformly 
distributed.

However, once the minimal value in the set is found and the relative bond 
broken,
the knowledge of this event makes the variables of the remaining 
non-broken bonds no longer simply uniformly distributed in the 
interval $[0,1]$, and correlated (no more independent one each other).
In fact, after the breaking of the bond with the minimal
threshold, one has to {\em condition} the probability of any further
event to the last known event.
This information influences the probability distribution of the remaining
bonds of the system and creates correlations among them \cite{Feller}.

The systematic study of this ``memory'' effect is what is called Run Time 
Statistics \cite{RTS,RTS-rig}.

In order to clarify the ``step by step'' mechanism of storage of 
conditional information, let us think to have fixed a time-order
path $A_t$, i.e. an {\em history} of the dynamics up to time $t$.
$A_t$ is given by the time ordered sequence 
$\{i_0, i_2, ... ,i_{t-1}\}$ of the broken bonds up to time $t$. 
Let us suppose to know the joint threshold probability density function
$P_t(\{x\}_{\partial C_t}| A_t)$ of the whole 
set of non-broken bonds conditioned 
to the knowledge of the past history $A_t$.
$P_t(\{x\}_{\partial C_t}|A_t)$ represents the 
``effective'' distribution of the disorder
at the $t^{th}$ time-step of a fixed history $A_t$.
Note that at time $t=0$, one has:
\begin{equation}
P_0(\{x\}_{\partial C_0})=\prod_{k\in S} p_0(x_k)=1\,,
\label{in-dist}
\end{equation}
where $S$ is the whole lattice, as no information from 
the dynamics is still present.

Since any kind of ``order'' relation, superimposed to a set of independent
stochastic variables, introduces correlations, in general 
$P_t(\{x\}_{\partial C_t}|A_t)$
does not factorize in the product of single-bond 
``effective'' density functions
for $t>0$ \cite{Feller}.
That is, it is not possible to write:
\begin{equation}
P_t(\{x\}_{\partial C_t}|A_t)=\prod_{k\in\partial C_t} p_{k,t}(x_k)\,.
\label{factor}
\end{equation}
However, as shown in \cite{RTS-corr}, in the limit of 
large number of variables 
the ``geometrical'' correlations in $P_t(\{x\}_{\partial C_t}|A_t)$ become
negligible, and one can make, at any time step, the approximation
given by Eq.~(\ref{factor}).
Therefore, we consider the approximated case where 
the ``effective'' probability 
density function of the disorder of the system, with all the 
information about the past history
stored, is given by the set of ``effective'' one-bond functions
$p_{k,t}(x)$ for each non-broken bond $k$.
The rigorous exposition of RTS, by using the non-factorizable function
$P_t(\{x\}_{\partial C_t}|A_t)$ at each $t$ is given in \cite{RTS-rig}.

Knowing the set of functions $p_{k,t}(x)$, one can write the ``effective''
probability $\mu_{i,t}$ that a given bond $i$ of the set 
is broken at that time.
It is simply the probability, conditioned to the whole past history, that  
$x_i$ is the minimum in the set of non-broken bond variables. 
Consequently, it is given by Eq.~(\ref{mu}), i.e.
\begin{equation}
\mu_{i,t}=\int_0^1 dx\,p_{i,t}(x) \left[ \prod_{k(\neq i)}^{\partial C_t}
\int_x^1 dy \, p_{k,t}(y) \right]\,.
\label{mu-app}
\end{equation}
The set of $\mu_{i,t}$ for each non-broken bond and for each time-step, 
defines a branching process of the dynamics; i.e. each history $A_t$ 
at time $t$ continues with a certain
probability $\mu_{i,t}$ in a different history $A_{t+1}$ at time $t+1$ 
for each breaking bond $i$ at time $t$.
In order to continue the probabilistic description of the branching at further 
time-steps, one should obtain the new
set of functions $p_{k,t+1}(x)$ for these 
different cases of breaking at time $t$,
using only the ``old'' set of $p_{k,t}(x)$ and the set 
of probabilities $\mu_{i,t}$
defining the branching.
This is possible by using the notions of conditional probability. 
Here the simple rule relating the conditional to joint probability of a 
first event $A$ to a second event $B$ is reminded \cite{Feller}:
\begin{equation}
\mbox{Prob}(A|B)=\frac{\mbox{Prob}(A\cap B)}{\mbox{Prob}(B)}\,,
\label{con-joint}
\end{equation} 
where, as usual, $A|B$ means the event $A$ {\em conditioned} 
to the event $B$, while
$A\cap B$ the event $A$ {\em joint} to the event $B$.

Note that ``memory'' up to time $t$, for a fixed 
history $A_{t}$ in the branching
of all the possible histories,
is already stored in the functions $p_{k,t}(x)$. Consequently, in
order to obtain the set of probability functions $p_{k,t+1}(x)$ for the history
$A_{t+1}$ obtained from $A_t$ adding the breaking of bond $i$ at time $t$,
one has to store only information about the last step.

At this point one has to distinguish the three cases: {\bf (1)} the just broken
bond $i$, {\bf (2)} a non-broken bond $j$ far from $i$, 
and {\bf (3)} a non-broken neighbor $l$ of $i$.\\
{\bf (1)} In this case let us call the conditioned probability density of bond
$i$ after its braking with $m_{i,t+1}(x)$ instead of $p_{i,t+1}(x)$, remarking
with this that after its breakdown, bond $i$ is removed 
definitely from the interface.
Note that, since after $t$ the bond $i$ does not participate to the dynamics,
its ``effective'' probability density will not change anymore. 
$m_{i,t}(x)dx$ is the probability that $x< x_i\le x+dx$, conditioned to the 
the past history up to its breaking. However, since the memory up to the 
time-step just before its breaking is stored in the known 
functions $p_{k,t}(x)$,
$m_{i,t}(x)dx$ is the probability, calculated using the set 
of functions$\{p_{k,t}(x)\}$, 
that $x< x_i\le x+dx$ (event $A$ of Eq.~(\ref{con-joint})) conditioned to the 
fact that the bond $x_i$ is the minimum in the set of 
interface bonds at that time (event $B$ of Eq.~(\ref{con-joint})).
Therefore, from Eq.~(\ref{con-joint}), one has Eq.~(\ref{m}):
\begin{equation}
m_{i,t+1}(x)=\frac{1}{\mu_{i,t}} p_{i,t}(x)\left[ 
\prod_{k(\neq i)}^{\partial C_t} \int_x^1 
dy\,p_{k,t}(y) \right]\, .
\label{m-app}
\end{equation}
In a quite similar way we can update the effective probability densities 
for the cases {\bf (2)} and {\bf (3)}.
In the case {\bf (2)}, using the set of functions $\{p_{k,t}(x)\}$,
$p_{j,t+1}(x)dx$ is the probability that $x<x_j\le x+dx$ (event $A$)
conditioned to the fact that $x_i$ was the minimal in the interface at 
time $t$. Again from Eq.~(\ref{con-joint}) one has Eq.~(\ref{p1}):
\begin{equation}
p_{j,t+1}(x)=\frac{1}{\mu_{i,t}} 
p_{j,t}(x)\!\!\int_{0}^{x}\!\! dy \,p_{i,t}(y) 
\!\!\left[ \prod_{k(\neq i,j)}^{\partial C_t}\!\!
\int_y^1 \!\!dz \, p_{k,t}(z) \right] \,.
\label{p1-app}
\end{equation}
In the case {\bf (3)} one has to distinguish the two different 
damaging rules, and the conditioning events are more complex.
For rule {\bf 1}, using the set of function $\{p_{k,t}(x)\}$, 
$p_{l,t+1}(x)dx$ is the probability that $x<x_l\le x+dx$ 
(event $A$) conditioned 
to the fact that $x_i$ was the minimum and that the value of $x_l$ 
at this time-step differs from the value at the previous time-step 
for a random fraction of itself (event $B$).
One, then, gets Eq.~(\ref{p21}):
\begin{eqnarray}
&&p_{j,t+1}(x)=\frac{1}
{\mu_{i,t}}\int_{0}^{1}dy\, \frac{1}{y} \theta(y-x)
p_{j,t}(y)\times\nonumber\\
&&\times\int_{0}^{y} dz \, p_{i,t}(z)
\left[ \prod_{k(\neq i,j)}^{\partial C_t}
\int_z^1 du \, p_{k,t}(u) \right]\,.
\label{p21-app}
\end{eqnarray}
Finally, for rule {\bf 2}, always using the set of functions $\{p_{k,t}(x)\}$,
$p_{l,t+1}(x)dx$  is the probability that $x<x_l\le x+dx$ 
(event $A$) conditioned 
to the fact that $x_i$ was the minimum and that the value of $x_l$ 
at this time-step differs from the value at the previous 
time-step for a fraction
$1/n_{i,t}$ of $x_i$ (event $B$). From this one has eq.~(\ref{p22}):
\begin{eqnarray}
&&p_{j,t+1}(x)=\nonumber \\
&&=\frac{1}{\mu_{i,t}}\!\int_0^1\!\! dy\!
\left[\!\prod_{k(\neq i,j)}^{\partial C_t}\!\int_y^1\!\! dz\, p_{k,t}(z)\right]
p_{i,t}(y)\,p_{j,t}\!\!\left(x+{y\over n_{i,t}}\right)\times
\nonumber \\
&&\times\theta\left(\frac{n_{i,t}}{n_{i,t}-1}x-y\right)
\theta\left(n_{i,t}(1-x)-y\right)\,.
\label{p22-app}
\end{eqnarray}

\begin{table}
\begin{centering}
\begin{tabular}{|c|c|c|c|} \hline
$     $ & $L=64$ & $L=128$ & $L=256$\\ \hline
$D_f$ (dam. rule {\bf1})  &$1.75(2)$  &$1.74(2)$ &$1.74(2)$\\ \hline
$D_f$ (dam. rule {\bf2})  &$1.73(2)$  &$1.75(2)$ &$1.76(2)$\\ \hline
\end{tabular}
\caption{Fractal dimension of the spanning cluster
 for different sizes and for the two damaging rules.}
\label{table1}
\end{centering}
\end{table}

\end{document}